\documentclass[times,authoryear,twocolumn]{elsarticle} %,twocolumn

\usepackage{jasr}
\usepackage{framed,multirow}

\usepackage{amssymb}
\usepackage{latexsym}

\usepackage[switch]{lineno}

\usepackage{url}
\usepackage{xcolor}
\definecolor{newcolor}{rgb}{.8,.349,.1}

\usepackage{graphicx} % for including images
\usepackage{subcaption}
\usepackage{amsmath} % for math formatting
\usepackage[utf8]{inputenc} % for handling special characters
\usepackage[T1]{fontenc} % for handling special characters
\usepackage{natbib}
\usepackage{siunitx}
\usepackage{tabularx}
\usepackage{booktabs}
\usepackage{tablefootnote} % for table footnotes
\usepackage{makecell}
\usepackage{float}

\PassOptionsToPackage{hyphens}{url}\usepackage{hyperref}
\usepackage{cleveref}
\usepackage{hhline}
\usepackage{multirow}
\usepackage{pgfplots}
\pgfplotsset{compat=1.18}

\newlength{\subcolumnwidth}

\newcommand{\nextsubcolumn}[1][]{%
  \cr\noalign{\hfill}
  \if\relax\detokenize{#1}\relax\else\hsize=#1\setlength{\subcolumnwidth}{\hsize}\fi
}

\journal{Advances in Space Research}

\begin{document}

\verso{Higgins \textit{et. al.}}

\begin{frontmatter}

\title{Predicting Cislunar Orbit Lifetimes from Initial Orbital Elements}
%Quick and Dirty Cislunar Orbit Lifetimes from Single Epoch TLEs via self-organizing Maps}

\author[1]{Denvir Higgins\corref{cor1}}
\cortext[cor1]{Corresponding author: 
  Email: higgins30@llnl.gov}

\author[1]{Travis Yeager}
\author[1]{Peter McGill}
\author[1]{James Buchanan}
\author[1]{Tara Grice}
\author[1]{Alexx Perloff}
\author[1]{Michael Schneider}

\address[1]{Lawrence Livermore National Laboratory,
  7000 East Ave,
  Livermore,
  94550,
  CA,
  USA}

\received{1 May 2013}
\finalform{10 May 2013}
\accepted{13 May 2013}
\availableonline{15 May 2013}
\communicated{S. Sarkar}

\begin{abstract}
%% context: cislunar space, why we care about lifetimes
\noindent Cislunar space is the volume between Earth's geosynchronous orbit and  beyond the Moon, including the lunar Lagrange points. Understanding the stability of orbits within this space is crucial for the successful planning and execution of space missions.
%% the problem: why it's hard? 
Orbits in cislunar space are influenced by the gravitational forces of the Sun, Earth, Moon, and other Solar System planets leading to typically unpredictable and chaotic behavior. It is therefore difficult to predict the stability of an orbit from a set of initial orbital elements.
%% this paper's solution: simulate & SOM
We simulate one million cislunar orbits and use a self-organizing map (SOM) to cluster the orbits into families based on how long they remain stable within the cislunar regime. Utilizing Lawrence Livermore National Laboratory's (LLNL) High Performance Computers (HPC) we develop a highly adaptable SOM capable of efficiently characterizing observations from individual events.
%% results: what do we find? numbers
We are able to predict the lifetime from the initial three line element (TLE) to within 10\% for 8\% of the test dataset, within 50\% for 43\% of the dataset, and within 100\% for 75\% of the dataset. The fractional absolute deviation peaks at 1 for all lifetimes. Multi-modal clustering in the SOM suggests that a variety of orbital morphologies have similar lifetimes. The trained SOMs use an average of 2.73 milliseconds of computational time to produce an orbital stability prediction.
%% implications of this work:
The outcomes of this research enhance our understanding of cislunar orbital dynamics and also provide insights for mission planning, enabling the rapid identification of stable orbital regions and pathways for future space exploration. As demonstrated in this study, an SOM can generate orbital lifetime estimates from minimal observational data, such as a single TLE, making it essential for early warning systems and large-scale sensor network operations.

\end{abstract}
\end{frontmatter}

\section{Introduction}
%cislunar space, why do we care
Cislunar space---the volume between Earth's geosynchronous orbit and beyond the Moon, including the lunar Lagrange points---is becoming increasingly important due to the growing interest in lunar exploration, satellite deployment, and space traffic management \citep[e.g.,][]{baker2024comprehensive}. This is driving advances in space domain awareness (SDA) techniques to ensure the safety and efficiency of cislunar operations \citep{Borowitz2023-tx}. Commercial cislunar interests include plans for combusting the hydrogen encased within the lunar surface with oxygen for efficient fuel production for space exploration \citep{fuel}. NASA's Lunar Gateway \citep{gateway} plans to create a communication network orbiting the Moon to support the Artemis mission \citep{Angelopoulos2014}. Alongside U.S. interests in cislunar space, the Outer Space Treaty \citep[e.g.,][]{kopal1966} ensures global participation in outer space exploration, including the exploration of cislunar space. Therefore, the number of satellites in cislunar space is likely to increase significantly over the coming decade.

%orbits in cislunar, why is it hard
%find existing literature on early warning systems, astronomical systems, distributed sensors
Critical to planning any cislunar activity or mission is an understanding of unaided cislunar orbits. Unaided orbits in cislunar space are difficult to characterize due to gravitational forces from the Sun, Moon, and planets all playing significant roles in this regime. This complicated force landscape leads to chaotic and typically unpredictable orbital dynamics \citep[e.g.,][]{folta2022astrodynamics}. This makes space traffic management in cislunar space challenging. For example, trying to accurately predict a satellite's unaided end-of-life trajectory in cislunar space is difficult \citep[e.g.,][]{jones2024comparison}. Understanding these leads to better end-of-life outcomes for objects in this space.

%current methods -> a lot more satellites, a lot more data
The lifetime of a particular orbit can be estimated through a full propagation of the trajectory far into the future and linked observations from proliferated sensors. Data streams from large surveys such as the Vera C. Rubin Observatory Legacy Survey of Space and Time \citep{Ivezic2019} render these techniques infeasible. LSST is going to detect satellites in at least 40,000 of its images over its 10-year survey \citep{hu2022satellite}. With at least one satellite in each of those images, a technique that can rapidly characterize objects in cislunar space based off of single observations to determine end-of-life outcomes and stability is necessary. The complexity of orbital mechanics, combined with the dynamic interactions of various celestial bodies, creates a rich dataset that can be effectively analyzed using machine-learning (ML) algorithms. For example, ML has been applied to create a constellation of sensors in cislunar space to detect and track objects \citep{constellationpaper} as well as Moon-based sensors to track families of orbits, which can inform mission planning and collision avoidance strategies \citep{moonsensorpaper}.

Among these algorithms, self-organizing maps (SOMs) stand out for their ability to visualize high-dimensional data and identify clusters without prior labeling. Recent studies have demonstrated the potential of SOMs to effectively analyze astronomical data. For instance, researchers have utilized SOMs to categorize estimate galaxy parameters \citep{latorre2024}, estimate photometric redshifts \citep{Masters_2015}, cluster X-ray sources \citep{refId0}, and sorting suvery images based on source morphology \citep{VANTYGHEM2024100824} revealing underlying patterns that may not be immediately apparent through conventional methods relying on human intervention \citep{Fotopoulou2024-zu}.

Determining orbit lifetime from data-sparse trajectories remains a critical issue for SDA applications. SDA sensors take measurements at stochastic time increments, necessitating rapid categorization based on few observations of orbits. This paper aims to explore the application of SOMs in determining the lifetime of orbits within cislunar space from initial Keplerian orbital elements. By leveraging SOMs, we enhance our understanding of orbital dynamics in this critical region, paving the way for more efficient early warning systems and enhancing space traffic management.

\section{Methods}
    \subsection{Simulating Orbits}
    To enable reproducibility for expert readers, we simulated one million orbits over six years in cislunar space using the Space Situational Awareness for Python package \citep[SSAPy;][]{ssapyprep}. The model incorporates the EGM2008 Earth gravity model \citep{earthmodel}, the GRGM1200A lunar gravity model \citep{lunarmodel}, solar gravity, solar radiation pressure, and a Harris--Priester atmospheric drag model. Orbit propagation was performed using a 7/8 Runge--Kutta integrator. Initial Keplerian orbital elements were uniformly sampled from the subsets listed in Table \ref{ssapyparams}, ensuring thorough coverage of the cislunar phase space.

    Each orbit in the dataset was assigned a lifetime, defined as the temporal duration it remains in cislunar space. Orbits were evaluated at hourly intervals, and termination occurred if any of the following criteria was met: the orbit passed within the geosynchronous radius (4.216 × \(10^5\) km), traveled beyond two lunar distances from Earth (7.688 × \(10^6\) km), or approached the Moon at a distance where impact is possible in the subsequent time step. Lunar impact potential is assessed based on whether the orbit's current velocity allows it to traverse the distance to the Moon within the next time step. This results in termination distances ranging from two to five lunar radii. This approach avoids integrating down to the lunar surface, significantly reducing computational requirements while ensuring the orbit's fate is definitively to impact the Moon.

        \begin{table} %H
            \caption{Simulation parameters for one million orbits propagated out to six years. Initialization parameters were uniformly sampled in a geocentric celestial reference frame (GCRF). EGM2008 and GRGM1200A models are from \citep{earthmodel} and \citep{lunarmodel}, respectively.} \label{ssapyparams}
            \begin{tabular}{p{45mm}llll}  
                \underline{Initialization}         \\
                Semi-major axis (a):               & GEO\tablefootnote{GEO: 4.216x$10^{5}$ km} to 2 LD\tablefootnote{Two Lunar Distances (2 LD): 7.688x$10^{6}$ km}  \\ 
                Eccentricity (e):               & 0 to 1   \\
                Inclination (i):               & 0 to $\frac{\pi}{2}$   \\
                True Anomaly (TA):              & 0 to $2\pi$   \\
                Argument of the Periapsis (PA):              & 0 to $2\pi$   \\
                Right Ascension of the Ascending Node (RAAN):            & 0 to $2\pi$   \\
                & \\
                \underline{Propagation}                  \\
                Integrator:								& 7/8 Runge-Kutta  \\
                Cross-sectional Area:				&		0.25 $m^{2}$ \\
                Mass:&		250 kg \\
                Drag coefficient:&		2.3 \\
                Radiation pressure coefficient:&		1.3 \\
                Timestep: 		& 1 hour \\
                & \\
                \underline{Gravity Models} 				\\
                Earth surface model:					& EGM2008 \\
                Lunar harmonics model:			& GRGM1200A \\
            \end{tabular}
        \end{table}
        % \footnotetext[1]
        % \footnotetext[2]

    \subsection{Self-Organizing Maps}
    A self-organizing map \citep[SOM;][]{kohonen} is an unsupervised ML technique that combines dimensionality reduction and clustering into a resulting 2-D map. The visual nature of the map facilitates human understanding of high-dimensional data. It organizes input data using synaptic connections (or cell weights) onto a lattice of postsynaptic neurons (or cells). The SOM algorithm consists of three stages: competition, collaboration, and adjustment \citep{kohonen2013essentials}.
    
    In the competition phase, each neuron competes to represent the input data. The winning neuron, referred to as the best matching unit (BMU), is the neuron that is most similar to the input data. It has the shortest euclidean distance to the input data in feature space. Here \textit{i} is the winning neuron, \textit{x} is the input data, \textit{j} is a neuron, and $w_{j}$ is the associated weight of the neuron:
        \begin{align} \label{comp}
		i &= argmin_{j}{||x-w_{j}||}_{2}\\
		j &\in [1,2,\ldots,m]\nonumber
	\end{align} 
    \quad In the collaboration phase, the winning neuron forms neighborhoods with adjacent neurons determined by a Gaussian function and a user defined initial neighborhood radius, $\sigma_0$. This radius is defined in Eq. \ref{sig}. The neighborhood threshold function $h_{ij}$, where $d_{ij}$ refers to the lateral distance between the winning neuron \textit{i} and neighbor neuron \textit{j}, is defined as,
	\begin{align} \label{rad}
		% h_{ij}(d_{ij}) &= e{^\frac{-d{^2}_{ij}}{2\sigma{^2}}}.
            h_{ij}(d_{ij}) &= e^{-d{^2}_{ij}/2\sigma{^2}}.
	\end{align}
    \quad The neighborhood is then defined as the set of cells $j$ for which $h_{ij}$ is over zero. The neurons in this neighborhood then adjust their weights as a function of \textit{n} iterations:
        \begin{align} \label{collab}
		w_{j}(n+1) &= w_{j}(n)+\mu(n)h_{ij}(n)[x-w_{j}(n)]
	\end{align} 
    \quad Eq. \ref{lr} and Eq. \ref{sig} define further parameters used to tune the SOM. The initial value for the learning rate, $\mu_0$, defines how a map node moves in the output space, given an input. The initial value for the neighborhood radius, $\sigma_0$, defines the size of a cluster on the output space, and is inputted into the aforementioned neighborhood function, Eq. \ref{collab}. Both of these are converging functions that asymptotically reach 0 at \textit{n} iterations:
	\begin{align} \label{lr}
		\mu &= \mu_0e^{n/-T}
	\end{align} 
	\begin{align} \label{sig}
		\sigma &= \sigma_0e^{n/-T}
	\end{align}
    \quad The constant \textit{T}, the time constant, depends on the initial values $\mu_0$ and $\sigma_0$ as well as the number of iterations. The side dimension of the map stays constant through the map construction process. The user is able to choose both side lengths, so square and rectangular maps are equally possible. The conventional starting point for number of cells \citep{Tian2014AnomalyDU}, where \textit{l} is the length of the input data, is defined as: 
 	\begin{align} \label{sidelength}
		m &= 5\sqrt{l}
	\end{align}
    The side length is given by taking the square root of \textit{m} for a square map. Choosing a side length larger than this can unintentionally bias the SOM results, as it has more choices for less data. To judge the quality of the output map, there are two metrics available: the topographic error and the quantization error \citep{Tu19}. The topographic error is a measure of how many data points have first and second BMUs in different neighborhoods. The quantization error measures the average difference between the weights of the BMU and the input data. Minimizing both of these errors yields a SOM that is most representative of the input data in feature space.
    
    We use the MiniSOM\footnotemark[3]\footnotetext[3]{https://github.com/JustGlowing/minisom} Python implementation for analysis in this paper \citep{vettigliminisom}. This SOM model is based on NumPy. The best SOM is chosen through a compromise of good sampling and high resolution, leveraging the topographic and quantization errors. We choose lifetime to be the label for the SOM. We adapted MiniSOM's accuracy to be a median absolute deviation between the test data's lifetime and the median lifetime of the cell the test data fell into. We test the SOM on a random subset of  20\% of the data.
    
    \subsection{Sampling}
\begin{figure}[] 
        \centering
        \includegraphics[width=0.9\columnwidth]{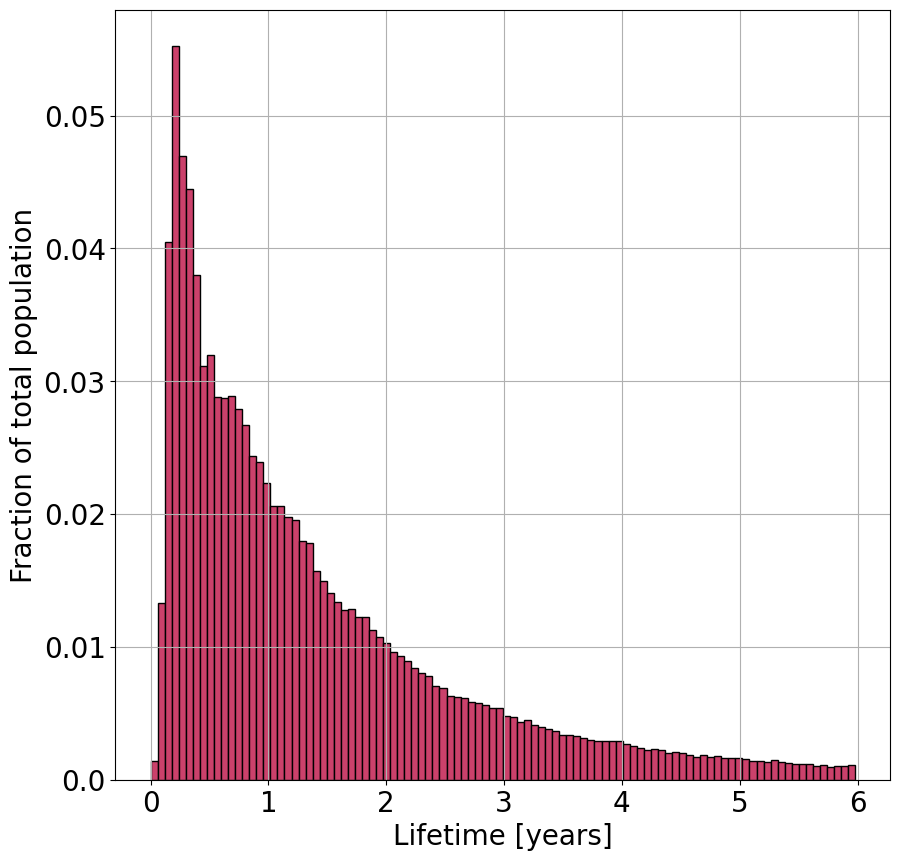}
        \caption{Lifetime distribution of simulated orbits with a lifetime in cislunar space between 1 and 6 years.}
        \label{datadistrib}
    \end{figure}

    We used the initial three line element (TLE) at timestep 1, consisting of six Keplerian orbital elements. We took a subset of 902,402 orbits constrained between 1 and 6 years lifetimes as our sample.  Fig. \ref{datadistrib} shows the distribution of lifetimes for orbits in our dataset.
    
    We then introduce this data to our pre-processing pipeline.
    
    \subsection{Feature Space Construction}
    We converted the orbit into a feature space that can be analyzed by the SOM. Each input has to have the same length and order of magnitude to be used effectively within a SOM. For each orbit, we first normalized the state vector. This is essential due to the order of magnitude differences between an instantaneous position and velocity. In this study, we tested both Z-Score and MinMax scaling normalization techniques using a scikit-learn \citep{scikitlearn} pre-processing package, which includes a MinMaxScaler and a StandardScaler.
    \quad The purpose of MinMax scaling is to maintain the original distribution of the data while making all of the values between 0 and 1. For each data point X, the minimal value of the dataset is subtract and then that is divided by difference between the minimal and maximal values. This is shown in Eq.\ref{mmnorm}.
    \begin{align} \label{mmnorm}
        X^{\text{minmax}}_{\text{normalized}} &= \frac{X-X_{\text{min}}}{X_{\text{max}}-X_{\text{min}}}
    \end{align} 
    When X is the minimal value, \(X^{\text{minmax}}_{\text{normalized}}\) returns 0. When X is the maximal value, \(X^{\text{minmax}}_{\text{normalized}}\) returns 1.
    \quad While MinMax scaling can handle varied distributions of data, z-score scaling assumes a Gaussian distribution. This method is also referred to as standardization as opposed to normalization. By using the mean $\mu$ and the variance $\sigma$ of the data, the normalized X is not restricted to a value range. This is demonstration in Eq. \ref{znorm}. 
    \begin{align} \label{znorm}
        X^{\text{z-score}}_{\text{normalized}} &= \frac{X-\mu}{\sigma}
    \end{align} 

    As a result, we had a lifetime label and normalized initial Keplerian orbital elements for each orbit in our dataset.
    
\section{Results}
    \subsection{SOM Construction}
\begin{figure}[]
          \centering
            \includegraphics[width=0.9\columnwidth]{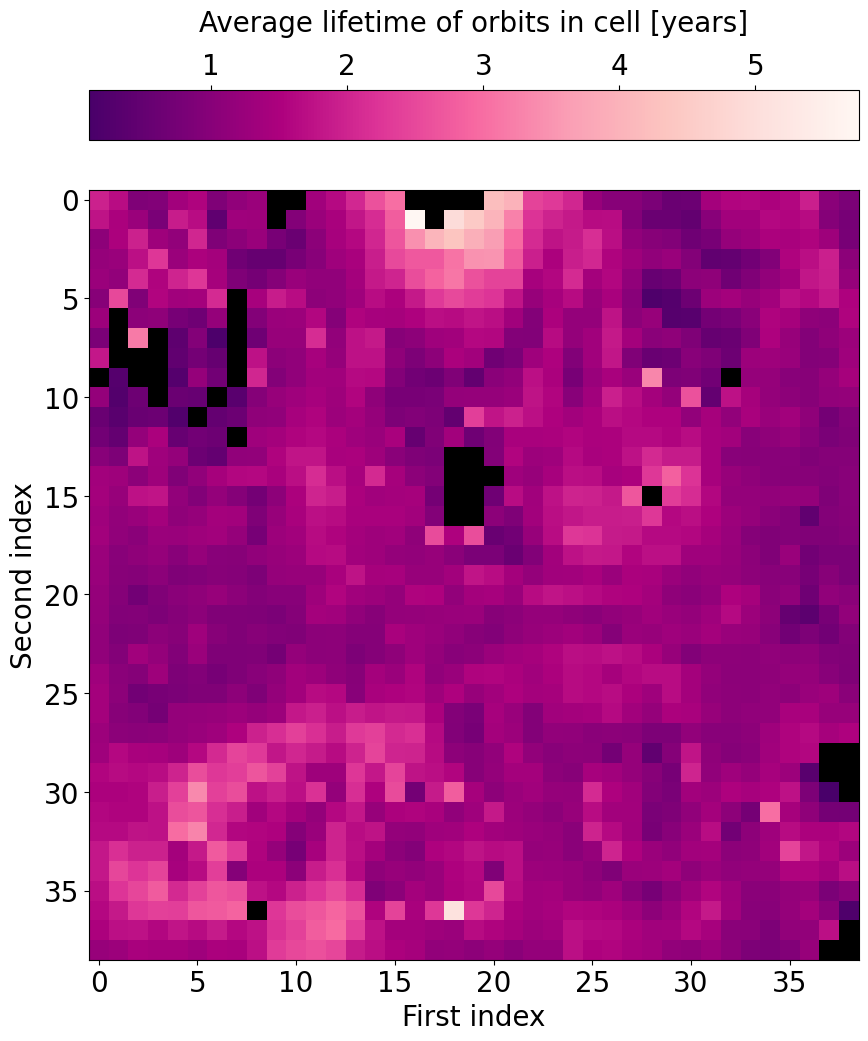} \\
        \caption{SOM generated from initial TLE, with a sequential purple color bar. The lighter cells correspond with cells with higher average lifetimes. The empty cells are denoted with black.} 
        \label{bestsom}
        \end{figure}

        The best performing SOM, with a quantization error of 0.321 and a topographic error of 0.163, was made using MinMax in its pre-processing with $\sigma_{0} = 10$, $\mu_{0} = 5$, and $n = 10^6$. This SOM is visualized in Fig. \ref{bestsom}. Note that the color map in this plot refers to the lifetime of the orbit, measured in years. The lighter the cell, the longer the orbit lasted, and vice versa for the darker cells. The SOM left some cells empty, which are colored in black.

        Density plots reveal how data clusters beyond the labels associated with the input. We observed rounded valleys of cells across the map containing fewer than 1,000 orbits, as well as single-cell peaks around the periphery of the map containing more than 5,000 orbits. These valleys strongly correlated with regions of high average lifetime, suggesting that the SOM effectively identified these orbits as distinct patterns within the dataset.
        
        \begin{figure}[]
          \centering
            \includegraphics[width=0.9\columnwidth]{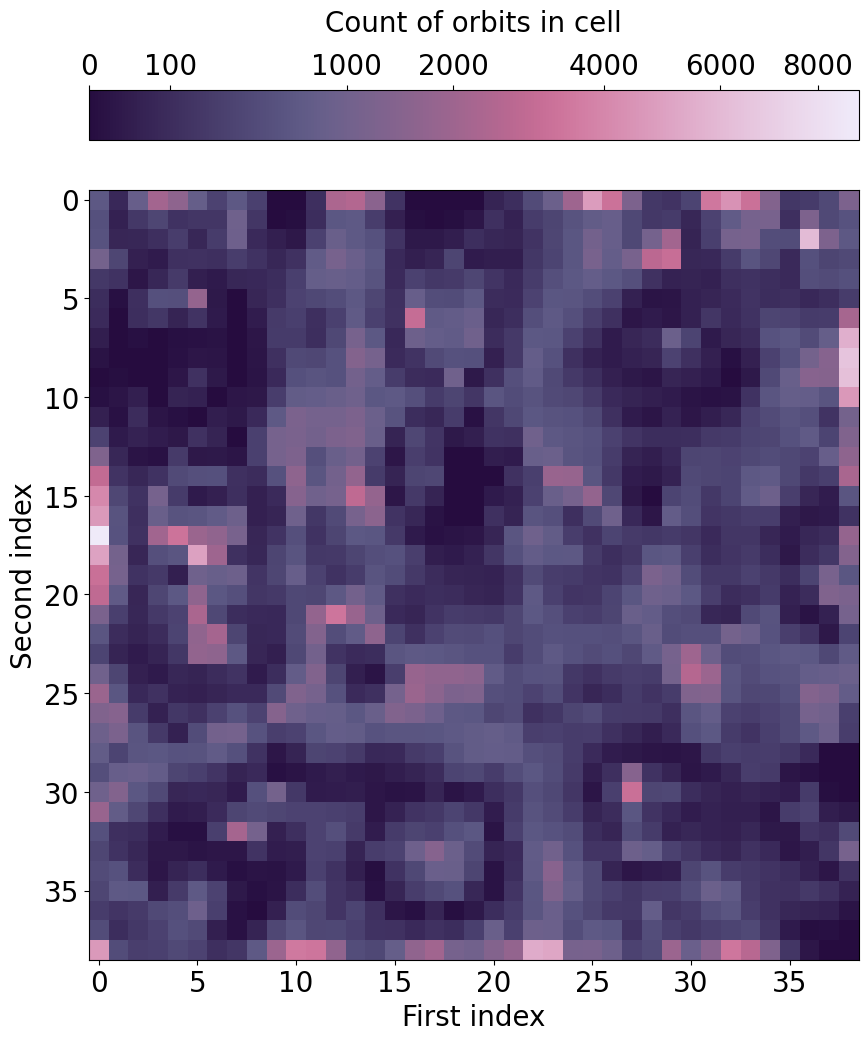} \\
        \caption{Density plot for SOM showing number of counts of orbits per cell. The colorbar \citep{Rollocmcrameri2024} is power law-scaled with $\gamma = 0.5$.} 
        \label{somfreq}
        \end{figure}

        The lifetime label variance within cells in the SOM is visualized to analyze the clustering. A high variance value indicates that orbits with contrasting lifetimes are sorted into the same cell, whereas a low variance indicates that orbits with similar lifetimes are grouped together. However, an orbit's lifetime is not the only parameter that determines whether orbits are physically similar, so high variance values within the SOM do not necessarily suggest ineffective clustering. A small minority of the data—only about 5.4\% of the orbits—has a lifetime exceeding 4 years. Therefore, it is expected that the variance in cells containing high-lifetime orbits will be greater than in the rest of the data.

        \begin{figure}[]
          \centering
            \includegraphics[width=0.9\columnwidth]{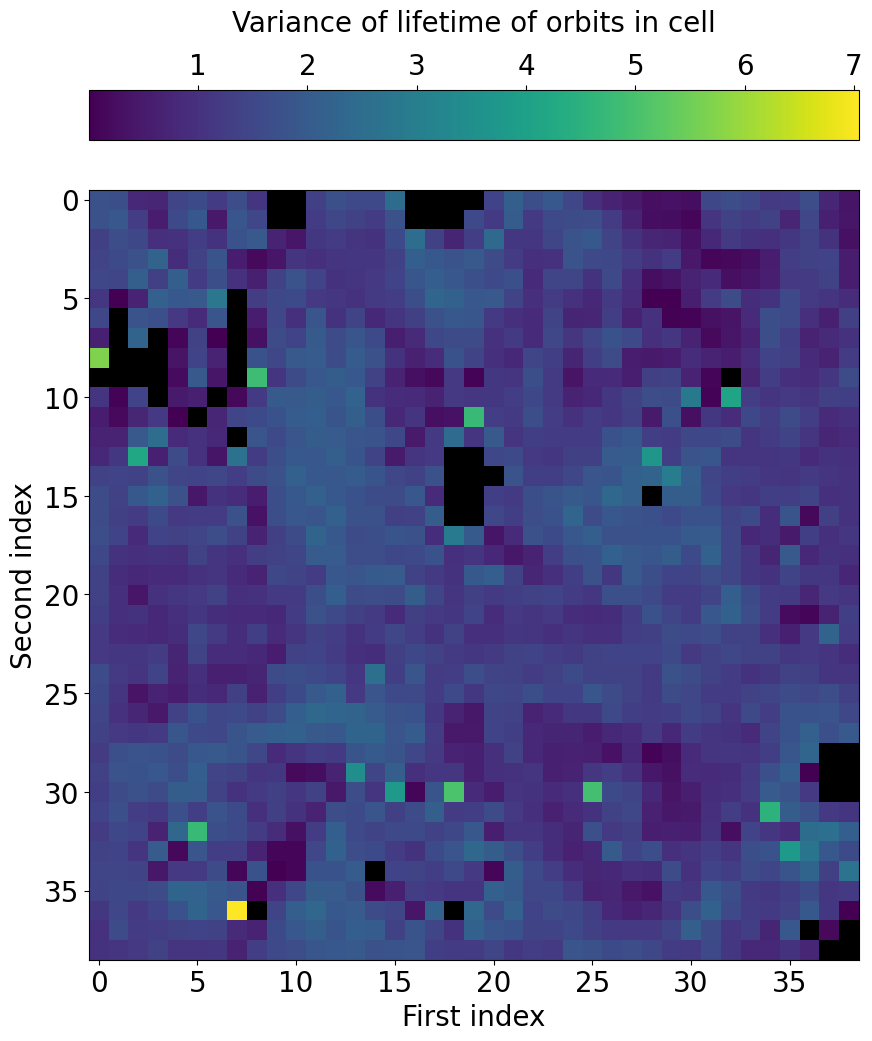} \\
        \caption{Variance of lifetime label in SOM. Cells containing single or zero orbits do not have an associated variance and are denoted with black.} 
        \label{somvar}
        \end{figure}

    \subsection{Lifetime Recovery}
    We tested the SOM on a $20\%$ subset of our data, 180,480 orbits. Each test orbit's lifetime recovery was computed in 2.72 milliseconds We took the absolute deviation between the raw lifetime label of each orbit with the median label of the cell it was sorted into. This is the median absolute deviation (MAD), which is explained further in Eq. \ref{absdevs}:

    \begin{align} \label{absdevs}
		MAD &= | L_{test} - median(L_{cell}) |
    \end{align}

    L represents the label associated with the test data. We used lifetime. 

    Additionally, we generate the fractional median absolute deviations, where the median absolute deviation is standardized by the test orbit's lifetime label. This accounts for the uneven distribution of orbit lifetimes.

    \begin{align} \label{fracabsdevs}
		MAD_{frac} &= \frac{| L_{test} - median(L_{cell}) |}{L_{test}}
    \end{align}

\begin{figure}[] 
        \centering
        \includegraphics[width=\columnwidth]{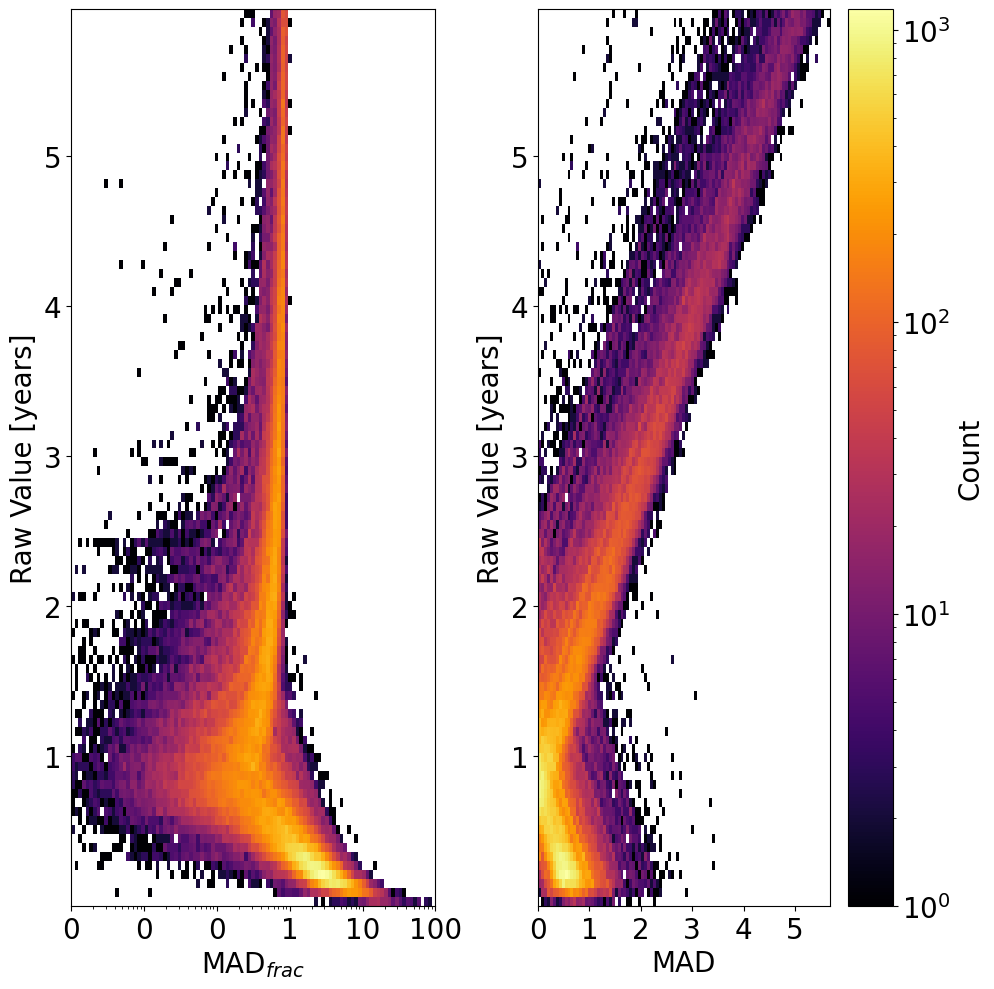}
        \caption{Density scatter plots of fractional and median absolute deviations versus lifetime values. The density of the region is represented by the colorbar. The range of lifetime recovery by the SOM at lifetimes less than 1 year is wider than for any other lifetime category, potentially due to the chaotic nature of these orbits.}
        \label{margfracdens}
    \end{figure}

    The relationships between test orbit lifetime and both $MAD$ and $MAD_{frac}$ are visualized in Fig. \ref{margfracdens}, showing a density plot showing a count of orbits in a region, with the deviation value on the x-axis and the raw lifetime value on the y-axis. 
    
\begin{figure}[] 
        \centering
        \includegraphics[width=\columnwidth]{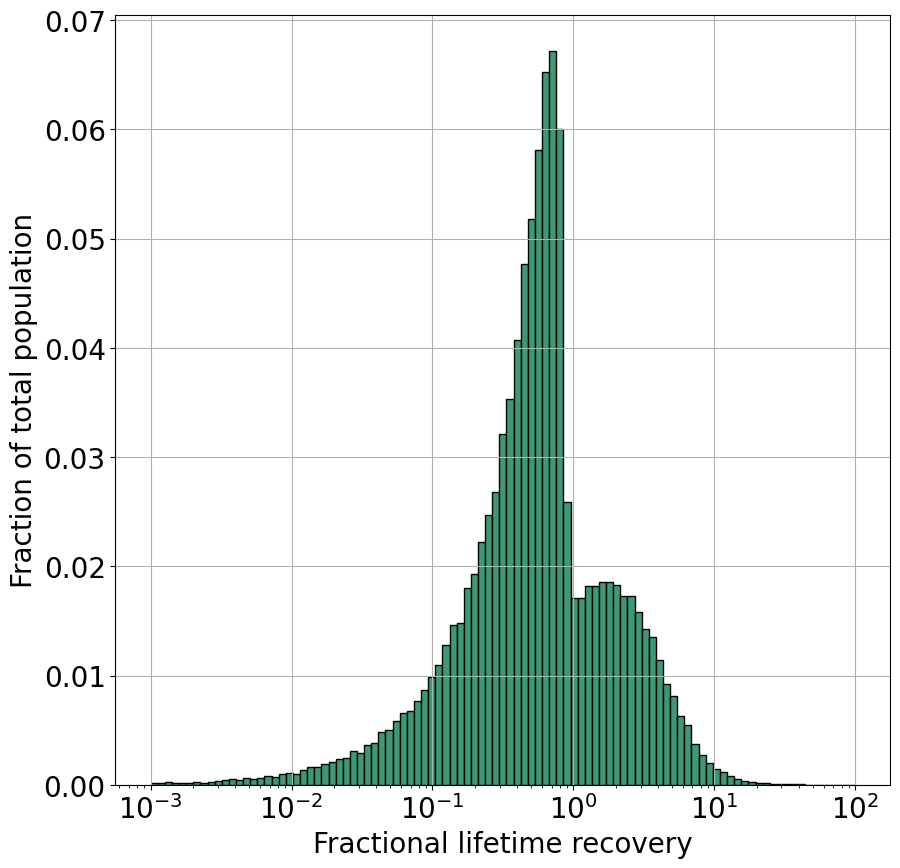}
        \caption{Count of fractional lifetime recovery with log-space binning. Out of the 180,480 test orbits, 15,905 orbits are recovered within 10\% of lifetime value, 78,467 within 50\%, and 135,673 within 100\%.}
        \label{fracrec}
    \end{figure}

    There was a prominence at 1 for fractional lifetime recovery in Fig. \ref{margfracdens}, suggesting that many orbits are recovered within 100\% of their lifetime value. For low lifetime values below 2 years, predominantly for those under 1 year, there was a positive skew to the fractional absolute deviations values, suggesting that the lifetimes these orbits are systematically overestimated. Orbits that survive less time in cislunar space are typically chaotic, and their initial TLE is not always deterministic of this chaos, so this overestimation was reflective of this nature of cislunar space. Further study on fractional median absolute deviation in Fig. \ref{fracrec} indicated that that majority of the 180,480 test orbits are recovered within 100\% of their lifetime, amounting to 135,673.  
    
\begin{figure}[] 
        \centering
        \includegraphics[width=\columnwidth]{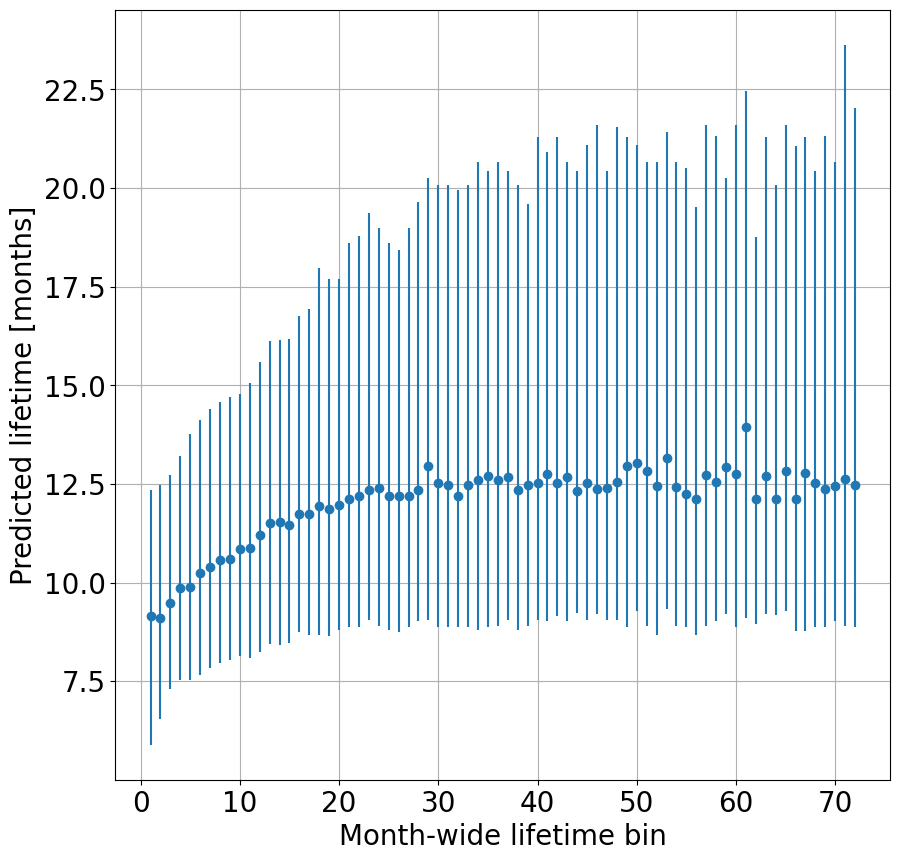}
        \caption{Relationship between lifetime of test orbit and predicted lifetime, with 68\% confidence interval error bars. Prediction increases in a logarithmic fashion as lifetime increases.}
        \label{binned}
    \end{figure}

    Indeed, 78,467 orbit's lifetimes were recovered within 50\%, and 15,905 were recovered within 10\%. This indicates that less than 25\% of the orbit's lifetimes were overestimated. The tail of the distribution represents orbits with lifetimes below 2 years, with the majority below 1 year. Beyond the test dataset as a whole, recovery as a function of lifetime was of interest. Monthly lifetime bins between 1 and 72 were created to test the SOM and recoup the associated lifetime recovery with a 68\% confidence interval, shown in Fig. \ref{binned}. As lifetime increased, the predicted lifetime value plateaued at 1 year and the confidence interval widened.

\section{Limitations of SOMs}
\begin{figure}[] 
        \centering
        \includegraphics[width=\columnwidth]{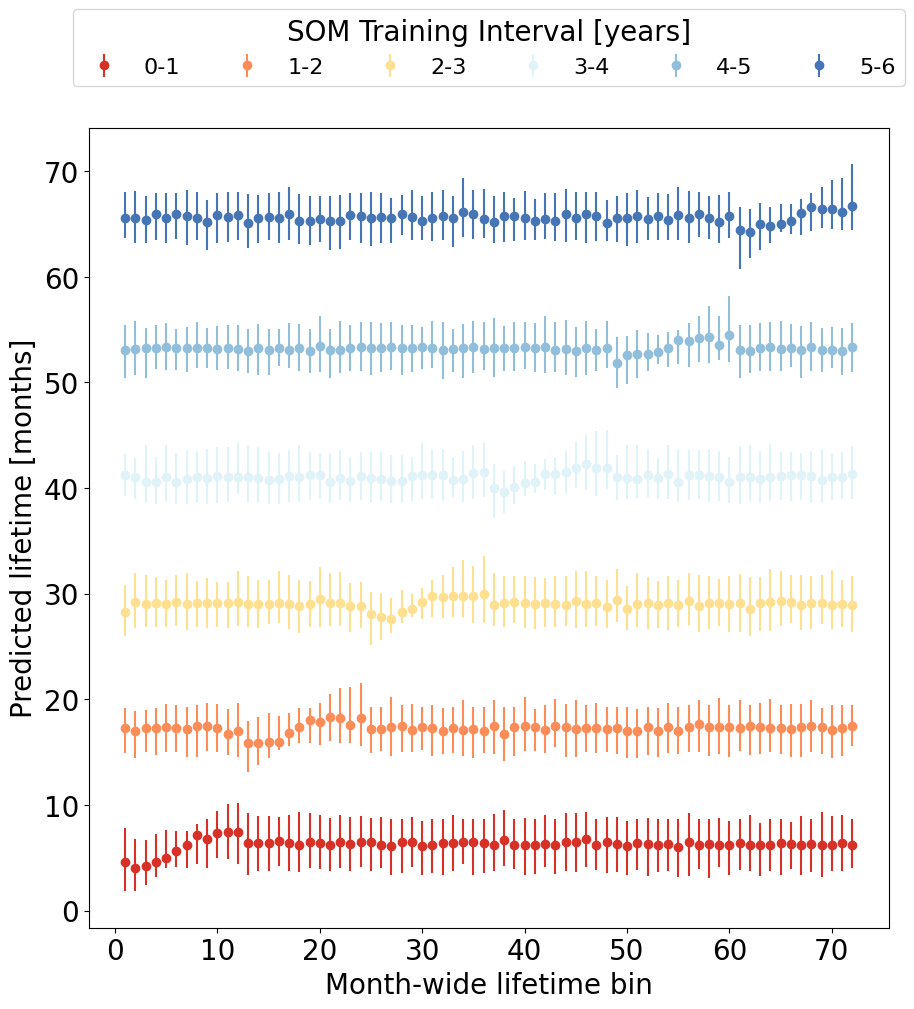}
        \caption{Relationship between lifetime of test orbit and predicted lifetime, with 68\% confidence interval error bars. Legend indicates the window of data the SOM was trained on.}
        \label{binnedmad}
    \end{figure}

We constructed six additional SOMs from evenly sampled data, ensuring that all lifetimes in month increments are equally represented. Each SOM was trained on a year window of data, then tested on data of all lifetimes. Fig. \ref{binnedmad} showcases the lifetime predictions for monthly windows of data for each pre-trained SOM. 

\begin{figure}[] 
        \centering
        \includegraphics[width=\columnwidth]{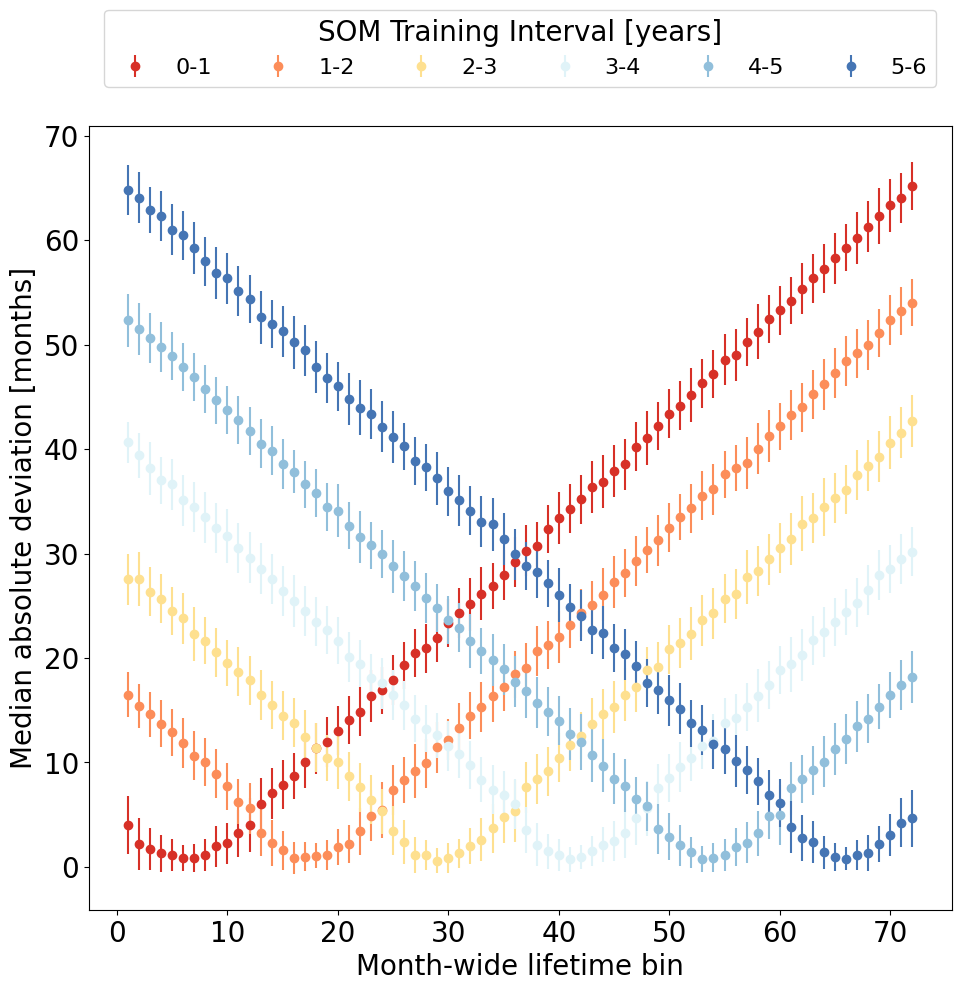}
        \caption{Median absolute deviation between lifetime of test orbit and predicted lifetime as a function of test orbit lifetime, with standard deviation error bars. Legend indicates the window of data the SOM was trained on.}
        \label{binnedpred}
    \end{figure}

The SOM accurately recognized the test data and predicted lifetimes within the training set window, recovering lifetimes of the test orbits within 2.385 months on average. A wave-like pattern was observed across the test dataset within the lifetime bin of the pre-trained SOM. For lifetimes outside the training window, there was no significant overestimation or underestimation which indicates that the SOM was ineffective at classifying data beyond this range. Figure \ref{binnedpred} reiterates \cref{binnedmad}, but this time through the median absolute deviation between the predicted lifetime and the true test orbit lifetime. The wave-like pattern observed in \cref{binnedmad} aligns with the minimum of each curve in \cref{binnedpred}. Lifetimes near the central value of each training set exhibit the minimal error in predicted lifetimes. The downward curve at the beginning of each training window, followed by an upward trend after the center point, indicates that the SOM is predicting lifetimes in the correct ``directions.'' This means that, for test data included in the training set, the SOM is statistically identifying the orbits correctly in a meaningful way.

The lowest MAD values for the test data corresponded with the window of data that the SOM was trained on. As a result of this study, we conclude that the SOMs are not useful outside their training sets. 
    
\section{Conclusion}
In this study, we aimed to classify unaided orbits in cislunar space using SOMs based on their initial TLE. This research demonstrated the potential of SOMs to enhance our understanding of orbital dynamics in the cislunar region, which is becoming increasingly important for future space exploration and satellite operations.

The results of our analysis revealed several key findings.
\begin{enumerate}
    \item The SOM provided a nearly instant guess toward whether an orbit would be stable for months or years for 75\% of the orbits based on their initial Keplerian orbital elements.
    \item The SOM produced lifetime estimations in an average of 2.72 milliseconds, leaving the door open for applications to real-time observing pipelines.
    \item The SOM included multi-modal lifetime distributions, with additional texture revealed through the density plot.
    \item The algorithm discerned patterns within the data that were independent of lifetime.
    \item The lifetime-restricted SOMs were ineffective when applied to data outside their training window, highlighting the limitations of SOMs in predicting anomalous test data.
    \item The SSAPy simulation produced a robust dataset of orbits exhibiting chaotic dynamics characteristic of this region.
\end{enumerate}

By characterizing cislunar orbits from single-event data, we enhance early warning systems and satellite deployment strategies, ultimately contributing to more efficient use of resources in space exploration. While the SOM was not highly specific in its results regarding lifetime, it demonstrated potential for identifying stable regions and families of orbits. The variability observed in results based on different sampling methods highlights the need for caution when generalizing findings.

Looking ahead, there are several promising directions for future research. Exploring alternative ML techniques could enhance real-time capability and improve classification accuracy. Longitudinal studies assessing the stability of classifications over time could further validate our findings and enhance their applicability. Of particular interest is the diffusion rate of cislunar families over time.

\section{Acknowledgements}
This work was performed under the auspices of the U.S. Department of Energy by Lawrence Livermore National Laboratory under Contract DE-AC52-07NA27344 and was supported by the LLNL LDRD Program under Project Number 22-ERD-054. The document number is LLNL-JRNL-872359.

\bibliographystyle{jasr-model5-names}
\biboptions{authoryear}
\bibliography{references}

\end{document}